\documentclass[prl,twocolumn,tightenlines,showpacs]{revtex4}%
\usepackage{amsfonts}
\usepackage{amsmath}
\usepackage{amssymb}
\usepackage{graphicx}
\begin{document}

\title{Detection of Geometric Phases in Flux Qubits with Coherent Pulses}% Force line breaks with \\

\author{Z.~H.~Peng}
\author{M.~J.~Zhang}
\author{D.~N.~Zheng}
\email[electronic address: ]{dzheng@ssc.iphy.ac.cn}
\affiliation{%
National Laboratory for Superconductivity, Institute of Physics and
Beijing National Laboratory for Condensed Matter Physics, Chinese
Academy of Sciences, Beijing 100080, PR China
}%
\date{\today}

\begin{abstract}
We propose a feasible experimentally scheme to demonstrate the
geometric phase in flux qubits by means of detuning coherent
microwave pulse techniques. Through measuring the probability of the
persistent current state in flux qubits, one can detect the Berry
phase that is acquired with system's Hamiltonian adiabatical
circular evolution in the parameter space. Furthermore, we show that
one could choose an appropriate amplitude of pulses in an experiment
to obtain high readout resolution when detuning frequency of pulses
is fixed and controlled phase shift gates can be implemented based
on the geometric phases by inductance coupling two flux qubits.
\end{abstract}

\pacs{Valid PACS appear here}

\maketitle

The unrivaled computational power of quantum computers in solving
certain problems that are virtually impossible for classical
computers has made quantum computing one of the hottest topics in
current research \cite{Nielsen}. The building blocks of quantum
computer are called quantum bits or qubits. A variety of different
physical systems have been explored to implement qubits. For
constructing a practical quantum computer, solid state qubits with
scability are believed to be very important. So far, superconducting
qubits based on Josephson junctions have emerged as a very promising
candidate of solid state qubits \cite{Makhlin,Yu,Mooij}. Up to now,
not only single superconducting qubits have been realized but also
entanglement between two coupled qubits has been observed
experimentally \cite{Izmalkov,Pashkin,Berkley}. Superconducting
qubits can be approximately divided into three categories, {\it
i.e.} charge qubit, flux qubit and phase qubit. Furthermore,
superconducting qubits, considered as artifical macroscopic
two-level atoms, are used to test fundamental laws in quantum
mechanics \cite{yangyu,Izmalkov,harmonic,Saito}.

When a quantum-mechanical system undergoes an adiabatic cyclic
evolution, it acquires a geometric phase factor, called Berry's
phase, in addition to the dynamical one \cite{Berry}. This effect
has been observed in many microscopic systems \cite{Atomita}. Based
on the geometric phase, one can realize any quantum logic operations
which can be composed from nontrivial two-qubit gates (such as a
controlled phase shift gate) and single-qubit gates
\cite{Lloyd,Wang}. Compared with conventional quantum computation,
the geometric quantum computation approach offers the potential of
fault-tolerance. The Berry phase does not depend on the details of
the motion along the path in parameter space but only depends on the
area enclosed by the system's parameters over a cyclic evolution in
parameter space. Therefore, the Berry phase is left unchanged by an
imperfection on the path as long as the area is left unchanged.
Because of this feature, it has been suggested that the Berry phase
could be a useful tool for intrinsically fault-tolerant quantum
computation \cite{Jones,Lelbfrled,Falci,Duan}. Conditional Berry
phase gates have been demonstrated in NMR and in trapped ions
\cite{Jones,Lelbfrled}. This is at the heart of geometric quantum
computation.

However, it is not clear whether the effect exists in macroscopic
quantum systems. Falci {\it et al}. \cite{Falci} have proposed a
scheme to observe the Berry phase and use the Berry phase to design
universal quantum logic gates in charge qubits. There are many other
schemes that have been proposed to detect geometric phase or realize
geometric quantum computation in superconducting qubits
\cite{Wang,Geometric}. However, almost all of these schemes are
based on charge qubits. In this letter, we show that these two
objects can also be realized in flux qubits with coherent pulse
techniques \cite{Abragam}. We believe that due to the long
decoherence time and relatively simple device fabrication process it
may be easier to observe the Berry phase in flux qubits than in
charge qubits \cite{Makhlin,Mooij,Falci}. Especially, dephasing time
of 4 Josephson junctions flux qubits have achieved 1$\mu s$
\cite{Nakamura}.

{\it Coherent pulse sequence technique.---}NMR techniques are often
used in coherent control quantum systems now being considered for
the implementation of quantum computer \cite{Vandersypen}. Collin
{\it et al}. \cite{Collin} have demonstrated NMR-like control of a
qubit superconducting circuit with multipulse sequences. More
recently, Kutsuzawa {\it et al}. \cite{Takayanagi2005} have observed
Ramsey fringes by applying a pair of phase-shifted microwave pulses
without introducing detuning. A clear advantage the method over
conventional detuning method is that it provides much faster
operation. The coherent pulse technique is one of the techniques
widely used in NMR \cite{Abragam}. The oscillating radio-frequency
field produced in a coil by a sequence of pulses occurring at times
$t_{1}, t_{2}, \ldots ,t_{k}$ and of durations $\tau_{1}, \tau_{2},
\ldots ,\tau_{k}$ can be represented by the function
\begin{equation}
\ V= \sum_{k}V_{k}(t)\cos({\omega_{rf}t+\varphi_{k})},
\end{equation}
where  $V_{k}(t)=0$ outside of the interval $t_{k} \leq t \leq
t_{k}+\tau_{k} $ and is approximately constant inside. The pulses
are called incoherent if the phases $\varphi_{k}$ are distributed at
random, and coherent if their values can be controlled. The
orientation of the rotating field with respect to the rotating frame
varies at random between pulses in the first method, but is well
defined in the second( in particular, can be made fixed ). Coherent
pulses have another advantage besides controlling the phase of the
rotating field: the signal-to-noise ratio is considerably improved.
Considering flux qubits, a circular motion of the system Hamiltonian
can be imposed by adiabatically varying the phase of the radio
frequency.

{\it Flux qubit and the rotational framework.---}We first consider a
single flux qubit and the measurement circuit is described in
Ref.\cite{Takayanagi2005,Chiorescu}. The relative phase shift
between these pulse sequences is precisely controlled by synchronous
microwave generators \cite{Takayanagi2005,Abragam}.

\begin{figure}
 \includegraphics[width=0.3\textwidth]{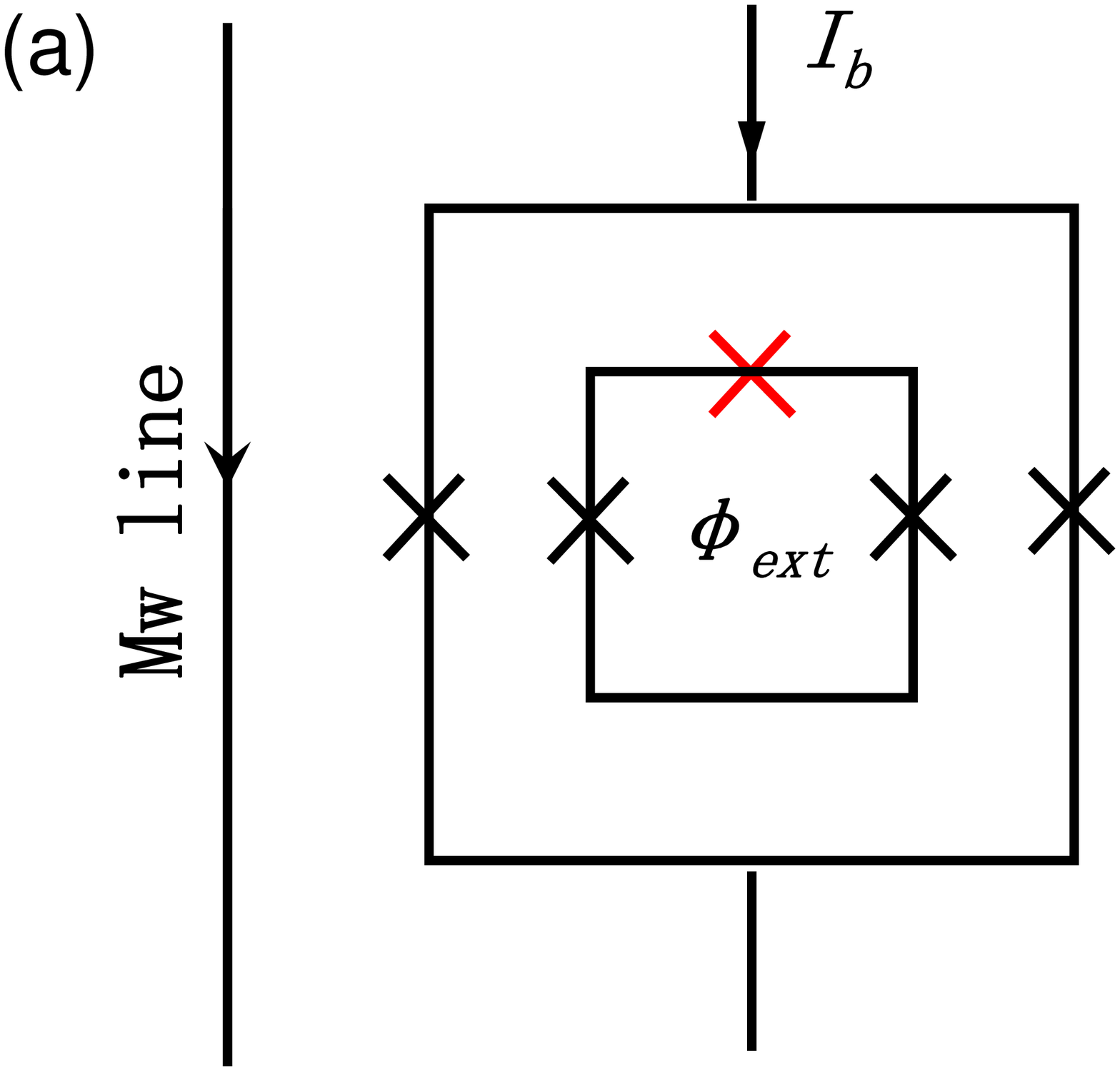}
 \includegraphics[width=0.3\textwidth]{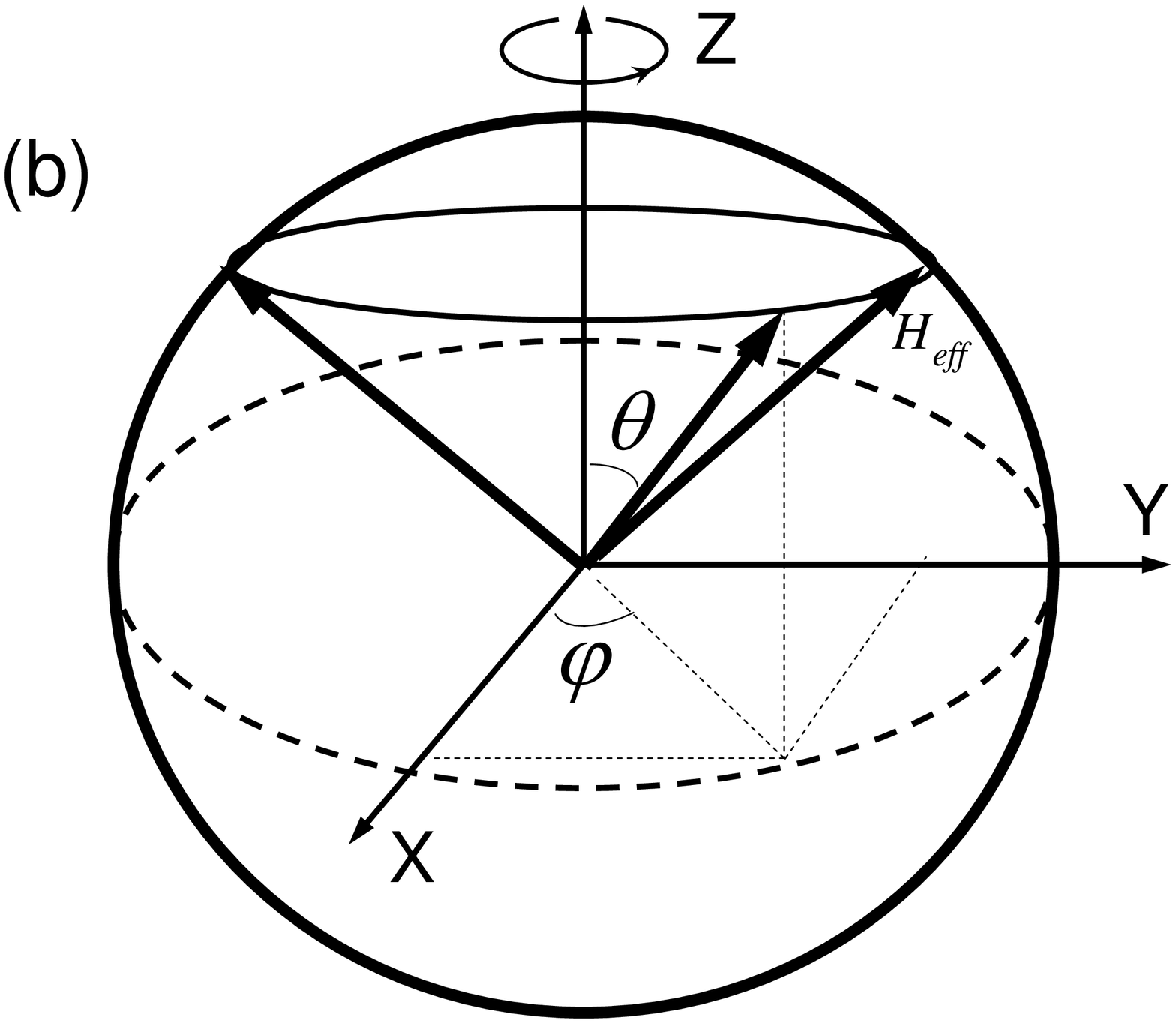}
\caption{ \label{fig:single} \footnotesize Schematic design of the
system. (a) The flux qubit, pierced by a magnetic flux $\phi_{ext}$,
and microwaves(MW) are applied to the qubit from the on-chip strip
line. The quantum state of the flux qubit can be read out by the
switching probability of the biased DC-SQUID. (b) Evolution of the
effective Hamiltonian in parameter space by adiabatically varying
the phase of the radio frequency. Each pulse makes the Hamiltonian
rotate about the $\it z$-axis at a fixed angle $\varphi$. After the
circular evolution, the Berry phase acquired in flux qubits is
$\pm\pi(1-\cos\theta)$, where $\varphi$ is the phase of the MW pulse
and $\theta=\arctan\frac{\nu\sin\eta}{\Delta\omega}$ is the cone
angle. }
\end{figure}

A flux qubit can be described as a two-state system or pseudo spin
1/2 particle \cite{Orlando}. In terms of the Pauli spin matrices
$\hat{\sigma}_{z}$ and $ \hat{\sigma}_{x}$, we can write the
Hamiltonian as $\hat{H_{0}}=\frac{\hbar}{2}(\varepsilon_{f}
\hat{\sigma}_{z}+\Delta \hat{\sigma}_{x})$, where $\Delta$ is tunnel
splitting, $\varepsilon_{f}\approx I_{p}(\phi_{ext}-1/2\phi_{0})$ is
the dc energy bias ($I_{p}$ is the circulating current of a flux
qubit). The two eigenstates of $\hat{\sigma}_{z}$ are
macroscopically distinct states with the qubit persistent current
circulating in opposite directions, i.e., the clockwise state
$|1\rangle$ and the counter-clockwise state $|0\rangle$. We employ a
microwave current burst to control the qubit state and induce
oscillating magnetic fields in the qubit loop. That event is
described by the perturbation Hamiltonian $\hat{V}=\frac{\hbar}{2}
V(t)\hat{\sigma}_{z}$, where $V(t)=2\nu\cos(\omega_{rf} t +
\varphi)$ has an amplitude correlated to the power of the applied
microwave pulses. The total Hamiltonian can be transformed by
rotating the matrix according to
$U(\eta)(\hat{H_{0}}+\hat{V})U^{-1}(\eta)$, where $U$ contains the
normalized eigenvectors of the Hamiltonian in the flux basis
\cite{Goorden},
\begin{equation}
U(\eta)=\left(\begin{array}{cc}
\cos\frac{\eta}{2} & -\sin\frac{\eta}{2} \\
\sin\frac{\eta}{2} & \cos\frac{\eta}{2}
\end{array} \right )
\end{equation}
where $\eta=\arctan(\Delta/\varepsilon_{f})$. We obtain the total
Hamiltonian
$\hat{H}=-\frac{\hbar}{2}\{[\omega_{0}-\frac{\varepsilon_{f}}{\omega_{0}}V(t)]\hat{\sigma}_{z}+\frac{\Delta}{\omega_{0}}V(t)\hat{\sigma}_{x}\}$,
where $\omega_{0}=\sqrt{{\varepsilon_{f}}^2+\Delta^2}$ is the qubit
Larmor frequency at the measured flux bias point. However, we can
disregard $\frac{\varepsilon_{f}}{\omega_{0}}V(t)$ because under the
usual experimental conditions,
$\frac{\varepsilon_{f}}{\omega_{0}}\nu<\omega_{0}$ \cite{Saito}.
This is analogous to the spin 1/2 particle, and because the Berry
phase is gauge-invariant, we write the Hamiltonian in the rotating
frame approxiamation \cite{Abragam}, $\hat{H}_{\rm
eff}=(\hbar\omega_{rf}/2)\hat{\sigma}_{z}+e^{i(\omega_{rf} t/2)
\hat{\sigma}_{z}}\hat{H}e^{-i(\omega_{rf} t/2) \hat{\sigma}_{z}}$
and obtain
\begin{equation}
\hat{H}_{\rm eff}=\frac{\hbar}{2} \left(
\begin{array}{cc}
\omega_{rf}-\omega_{0} & -\nu e^{i{\varphi}} \sin{\eta}\\
-\nu e^{-i{\varphi}} \sin{\eta} & \omega_{0}-\omega_{rf}\\
\end{array}
\right)
\end{equation}

The effective Hamiltonian is read $\hat{H}_{\rm
eff}=-(1/2)B\cdot\hat{\sigma}$, so we have defined fictitious field
$B\equiv\hbar(\nu\cos\varphi \sin\eta,\nu\sin\varphi \sin
\eta,\Delta \omega)$ , where $\Delta \omega=\omega_{0}-\omega_{rf}$.
The flux qubit thus behaves like a spin 1/2 particle in a magnetic
field. Under the action of coherent pulse sequences, the effective
qubit Hamiltonian describes a cylindroid in the parameter space
$\textit \{B\}$.

{\it Creating the Berry phase in a single flux qubit.---}Analogous
to a spin 1/2 particle, the Hamiltonian lies close to the $\it
z$-axis when $\left| \nu \right|\ll \left|\Delta \omega\right|$,
while the Hamiltonian lies close to the $\it x-y$ plane when $\left|
\nu \right|\gg \left|\Delta \omega\right|$. If the applied
radio-frequency radiation is far from resonance, the Hamiltonian is
quantized along the $\it z$-axis, and if the radio frequency is
swept towards resonance ($\Delta \omega=0$ ), the Hamiltonian
rotates from the z-axis towards the $\it x-y$ plane. If the
resonance is approached sufficiently slowly, the spin will follow
the Hamiltonian in the rotating frame according to the adiabatic
theorem. Next, by adiabatically varying the phase of the radio
frequency, a circular motion can be performed. When the Hamiltonian
returns to the $\it x-y$ plane the frequency sweep may be reversed,
so that the spin returns to its original state, aligned along the
$\it z$-axis. The Berry phase acquired in this cyclic process is
$\pm \pi$, where the signs $\pm$ depend on whether the system is in
the eigenstate aligned with or against the Hamiltonian. If we don't
sweep the radio-frequency all the way to resonance, but only to some
final value $\omega_{rf}$, the Hamiltonian ends at some angle to the
$\it z$-axis, and so flux qubit with arbitrary cone angles can be
realized. A similar case occurs if one replaces the frequency sweep
with an amplitude sweep, in which the radio frequency is always
applied in off-resonant conditions, and its amplitude is raised from
zero to some final value, $\nu$.

\begin{figure}
\includegraphics[width=0.4\textwidth,height=0.2\textheight]{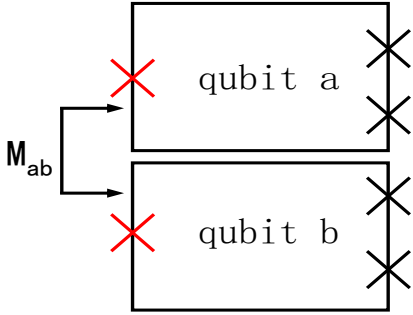}
\caption{\label{fig:coupled} \footnotesize Schematic design of the
two-bit gate. The two qubits can be coupled via their mutual
inductance. The Berry phase of the target qubit (qubit a) can be
controlled by the control qubit (qubit b).}
\end{figure}

We now describe a procedure to measure the Berry phase in
experiments. First, the flux qubit is prepared in the ground state
of the Hamiltonian at the degeneracy point and then a hard pulse is
applied to excite the flux qubit in a linear superposition of the
energy eigenstates which is analogous to the splitting of the photon
wavefunction at the first beamsplitter of a Mach-Zender
interferometer. Next, appropriate amplitude adiabatic coherent pulse
sequences are applied. With the geometric phases acquired in this
way, we will have an additional dynamical component, which depends
on experimental details. This can be eliminated using a conventional
spin echo approach: the pulse sequence is applied twice, with the
second application surrounded by a pair of $180^{0}$ pulses applied
to the spin particle. The final step in the procedure is to measure
the persistent current state of the qubit. The probability of
measuring the persistent current state $|1\rangle$ in the flux qubit
at the end of this procedure is given by
\begin{equation}
\ P(1)=\sin^{2}(2\gamma),
\end{equation}
independent of the elapsed time, where $\gamma$ is the Berry phase
that is acquired in the first adiabatic loop process. If the initial
state of the flux qubit is not prepared at the degeneracy point, we
have to deduce the Berry phase value $\gamma$ from the interference
pattern of the persistent current state $|1\rangle$ measuring
probability.

For the quantum mechanics system, the acquired geometric phase is
equal to half of the solid angle subtended by the area in the
parameter space enclosed by the closed evolution loop of fictitious
magnetic field,
\begin{equation}
\gamma=\pi(1-\frac{\Delta \omega}{\sqrt{\Delta
\omega^{2}+(\nu\sin\eta)^{2}}})
\end{equation}
If we fix the bias point at the degeneracy point, $\sin\eta=1$,
geometric phases to achieve a NOT gate for $\nu=\sqrt{3}\Delta
\omega $ and Hadamard gate for $\nu=\frac{\sqrt{7}}{3} \Delta
\omega$.

{\it The adiabatic conditional geometric phase gate.---}For the
implementation of a universal two-qubit gate, we consider in a
system of two coupled flux qubits with their mutual inductance. The
two qubits are now described by the total Hamiltonian:
\begin{equation}
\hat{H}={\textstyle \sum_{i=a,b}} \hat{H}_{i}+\it J
{\hat{\sigma}_{z}}^{(a)} {\hat{\sigma}_{z}}^{(b)}
\end{equation}
Two qubits can be coupled via the magnetic inductance or a large
Josephson junction for different coupled strength \cite{Haar}.

\begin{figure}
 \includegraphics[width=0.4\textwidth]{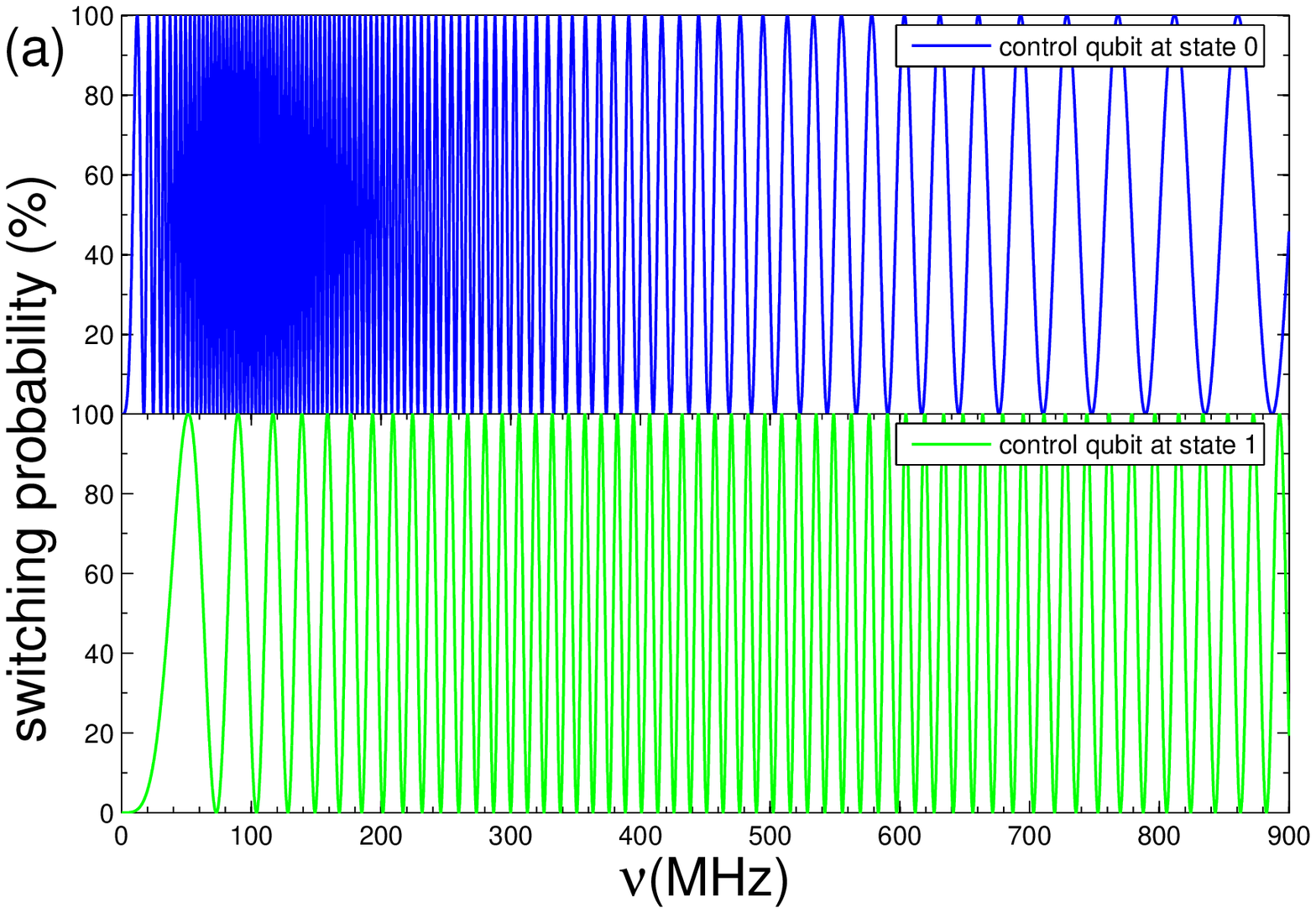}
 \includegraphics[width=0.4\textwidth]{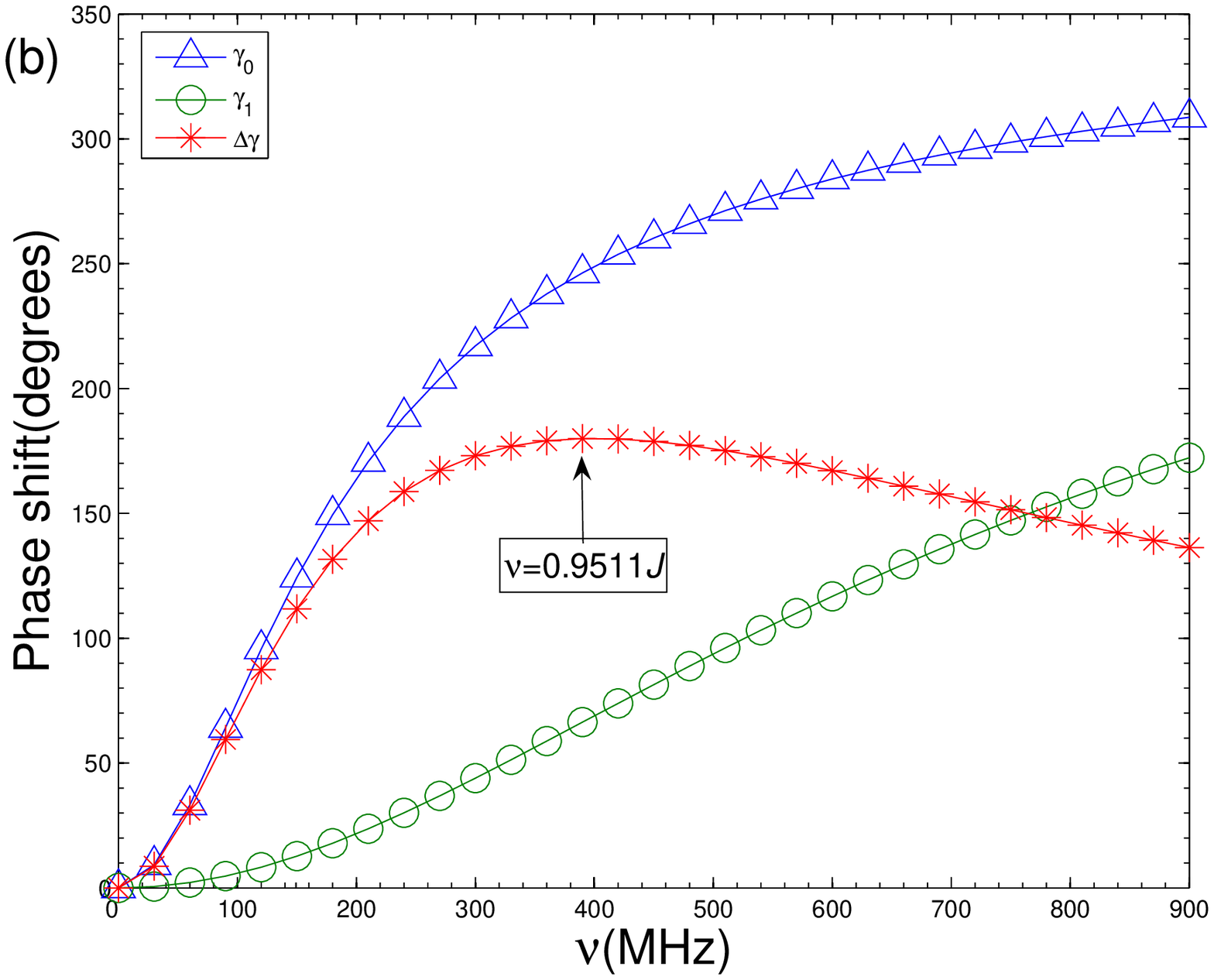}
\caption{\label{fig:phase} \footnotesize Calculated results of the
controlled phase shift in the target qubit. Here, the pulse sequence
is applied twice in order to eliminate the dynamics phase. The split
energy of the two coupled flux qubits is chosen from
Ref.\cite{Izmalkov}, $\textit J$=420MHz, $\Delta \omega$=129.78MHz.
(a) The switching probability of the persistent current state
$|1\rangle$ in target qubit (qubit a) vs. amplitude of coherent
pulses. At low amplitude region, the switching probability of the
target qubit when the control qubit at state $|0\rangle$ oscillates
very quickly. Therefore, one should choose an appropriate amplitude
of pulses in an experiment to obtain high readout resolution. (b)
Berry phase $\gamma_{0}$, $\gamma_{1}$ and the controlled Berry
phase difference $\Delta \gamma$ as a function of the amplitude of
microwave pulse. Open triangles, circles and stars denote
$\gamma_{1}$, $\gamma_{0}$ and $\Delta \gamma$ respectively. Near
the $\nu=0.9511J$, controlled phase shift dependence on the
$\Delta\omega$ is reduced to second order. }
\end{figure}

We consider that the $\Delta$ of the two qubits is different and the
Larmor frequency of the two qubits is different at the degeneracy
point, and only {\it qubit a} is close to resonance. In practice,
the two qubits usually have different parameter values because of
fabrication limitations \cite{Majer}. The two transitions of {\it
qubit a} (corresponding the two current states of {\it qubit b})
will be split by $\pm\pi \it J$, and so will have different
resonance offsets. With the coherent radio-frequency microwave
pulses, the effective Hamiltonian depends on the resonance offset,
and so the cone angle (and hence the Berry phase acquired) will
depend on the state of {\it qubit a}. This permits a conditional
Berry phase to be applied to {\it qubit a}, where the size of the
phase shift is controlled by {\it qubit b}. If the frequency is
applied at a frequency $\Delta \omega$ away from the resonance
frequency of {\it qubit a} when {\it qubit b} (the control qubit) is
in state 0, and $\nu$ is the amplitude of radio-frequency pulses,
then the differential Berry phase shift
\begin{equation}
\Delta \gamma=\pm \pi(\frac{\Delta \omega+J}{\sqrt{(\Delta
\omega+J)^{2}+(\nu\sin\eta)^{2}}}-\frac{\Delta
\omega}{\sqrt{\Delta \omega^{2}+(\nu\sin\eta)^{2}}})
\end{equation}
depends only on $\Delta \omega$, $\nu$ and $\textit J$; it is
independent of how the process is carried out as it is slow enough
to be adiabatic, but rapid compared with the decoherence times. A
range of controlled Berry Phases can be obtained by choosing
appropriate values of $\Delta \omega$ and $\nu$ ($\textit J$ is
fixed by the two coupled flux qubit with mutual inductance.) If we
fix the value of $\Delta\omega/J$, the controlled Berry phase will
rise and then fall as $\nu$ is increased. We can make the approach
very robust when the desired $\Delta\gamma$ occurs at the maximum in
this curve. Then the dependence on the $\Delta\omega$ is reduced to
second order \cite{Jones}. To achieve the controlled-NOT gate, we
fix the bias point at the degeneracy point $\sin\eta=1$ and we can
design a controlled $\pi$ shift at $\Delta\omega=0.309J$ and
$\nu=0.9511J$, as showed in Fig~\ref{fig:phase}.

We desicribe an example to illustrate the detection of the Berry
phase in a single flux qubit by applying coherent pulse sequences.
The parameters were as follows, split energy $\Delta=6$GHz, detuning
frequency $\Delta\omega$=150MHz, phase sweeps implemented using 100
linear steps of about 3ns, giving a total pulse sequence length of
about 600ns. These parameters satisfy the adiabatic threshold
$\frac{\nu}{T_{2}}\ll \frac{dH_{eff}}{dt}\ \ll \nu^{2}$
\cite{Abragam}, where $T_{2}$ is the dephasing time of the flux
qubit. The required adiabatic manipulation time not longer than
current flux qubits' performance and the required coherent pulse
techniques is mature \cite{Abragam,Vandersypen,Nakamura}. These
indicate we can detect the geometric phase in a flux qubit with
coherent pulses in experiments.

{\it Conclusion.---}We have proposed a new experimental scheme to
detect geometric phases with coherent microwave pulses in flux
qubits. Through measuring the persistent current state in the flux
qubit, we can detect the Berry phase which is acquired in circular
motion. Furthermore, we have designed one-bit quantum logic gate and
two-bit controlled phase shift gate based on the Berry phase. The
idea of the geometric phase shift gate demonstrated by the flux
qubits system here should in principle also work for other
superconducting qubits.

\begin{acknowledgments}
We thank P. L. Lang, A. Maassen van den Brink, C. P. Sun, W. M. Liu
and A. Ahmad for useful discussions and J. Q . You for useful
help.This work is supported by National Nature Science Foundation of
China(10221002), Ministry of Science and Technology of China through
the 973 program(2006CB601007) and Chinese Academy of Sciences.
\end{acknowledgments}

\end{document}